\documentclass[aps,prl,twocolumn,groupedaddress,showpacs]{revtex4}

\usepackage{graphicx}
\usepackage{amsmath}

\begin{document}

\title{Scaling relation for the superfluid density in cuprates - origins and limits}

\author{J.L. Tallon$^1$, J.R. Cooper$^2$, S.H. Naqib$^{1,2}$ and J.W. Loram$^2$}

\affiliation{$^1$MacDiarmid Institute for Advanced Materials and
Nanotechnology, Industrial Research Ltd., and Victoria University,
P.O. Box 31310, Lower Hutt, New Zealand.}

\affiliation{$^2$Interdisciplinary Research Center in
Superconductivity, Cambridge University, Cambridge CB3 0HE,
England.}

\date{\today}

\begin{abstract}
A universal scaling relation, $\rho_s \propto \sigma(T_c)\times
T_c$ has been reported by Homes $et$ $al$. (Nature (London) {\bf
430}, 539 (2004)) where $\rho_s$ is the superfluid density and
$\sigma(T)$ is the DC conductivity. The relation was shown to
apply to both $c$-axis and in-plane dynamics for high-$T_c$
superconductors as well as to the more conventional
superconductors Nb and Pb, suggesting common physics in these
systems. We show quantitatively that the scaling behavior has
several possible origins including, marginal Fermi-liquid
behavior, Josephson coupling, dirty-limit superconductivity and
unitary impurity scattering for a $d$-wave order parameter.
However, the relation breaks down seriously in overdoped cuprates,
and possibly even at lower doping.
\end{abstract}

\pacs{74.25.Gz, 74.25.Fy, 74.25.Ha, 74.25.Bt}

%%\begin{multicols}{2}
\maketitle

The absence of a recognized theory of high-$T_c$ superconductors
(HTS) has led to a search for universal relationships that might
provide a guide to the essential physics for HTS. The Uemura
relation\cite{Uemura1,Uemura2} is one such scaling relation,
namely $\rho_s(0) \propto T_c$, where $T_c$ is the superconducting
(SC) transition temperature and $\rho_s(0) = \lambda_{ab}^{-2} =
\mu_0 e^2 n_s/m^*$ is the superfluid density. Here $n_s$ is the
density of SC electrons, $m^*$ is their effective mass and
$\lambda_{ab}$ is the in-plane London penetration depth. The
Uemura relation has been invoked in support of Bose-Einstein
condensation of real-space pairs\cite{Uemura2,Dzhumanov} and is
generally discussed as a test of theoretical models\cite{Lee}.
However, there is increasing recognition that this relation is an
oversimplification\cite{Tallon1,Pereg}, and it breaks down on the
overdoped side of the SC phase curve\cite{Niedermayer}.

Recently, a new scaling relation, $\rho_s \propto \sigma(T_c)
\times T_c$ was reported by Homes $et$ $al.$\cite{Homes} where
$\sigma(T)$ is the DC conductivity. This was shown to apply over
five orders of magnitude including both $c$-axis and in-plane
dynamics for HTS as well as to the conventional superconductors Nb
and Pb. The authors suggested this relation may provide new
insights into the origins of SC in HTS materials. We examine this
relation and show that it is a natural consequence of several
quite different but well-understood mechanisms, including
dirty-limit conventional SC (Pb and Nb), Josephson coupling along
the $c$-axis, marginal Fermi-liquid behavior (optimal and
underdoped HTS), in-plane Josephson-coupled granular SC (strongly
underdoped HTS) and Abrikosov-Gor'kov $d$-wave pair-breaking for
non-magnetic scatterers. Since we submitted this work Homes $et$
$al.$\cite{Homes1} have reported some overlapping results but here
we examine the doping and impurity dependence of this relation
showing that, in overdoped HTS, it breaks down seriously. It is
perhaps significant that, here, the cuprates progress towards more
conventional SC and normal state (NS) behavior.

Because $\lambda^{-2} = \mu_0 e^2 n_s/m^*$ and $\sigma =
ne^2\tau/m^*$ the correlation is equivalent to a relationship
between scattering rate 1/$\tau$ and $T_c$ given by $\hbar/\tau =
2.7\pm0.5 \times k_BT_c$. Here we assume that all the spectral
weight associated with the free carriers condenses into the
$\delta$-function at $\omega$=0, i.e. $n_s$=$n$. In fact, for
underdoped cuprates some spectral weight remains at finite
frequency\cite{Tanner}. However, we assume that this weight is
related to some other excitation, different from the free
carriers, which does not contribute to $\sigma$($\omega$=0),
$\rho_s$ or the low-$T$ specific heat.

In considering possible origins for this relation between
$\hbar/\tau$ and $k_BT_c$ we note that it is a direct prediction
from Marginal-Fermi-liquid theory, where\cite{Varma,Timusk}
\begin{equation}
\ \hbar/\tau \approx \pi k_BT + \hbar\omega.
\end{equation}
At low frequency, the scaling relation is almost exactly recovered
provided $\tau$ is evaluated at $T_c$, as required.

That Nb and Pb fit the scaling line is surprising, but it is
readily shown that for classical SC in the dirty limit $\rho_s$
is, again, proportional to $\sigma(T_c) \times T_c$. For a
mean-free path, $\ell$, and a BCS coherence length, $\xi_0 = \hbar
v_F/(\pi\Delta)$, the effective penetration depth is given
by\cite{Tinkham}
\begin{equation}
\ \lambda_{eff} = \lambda_L (1 + \xi_0/\ell)^{1/2} \approx
\lambda_L (\xi_0/\ell)^{1/2} .
\end{equation}
Taking $\ell = v_F \tau$ and $\Delta = 1.76 k_BT_c$, we find that
$\rho_s$ is a factor of two higher than the scaling line
($\hbar/\tau = 5.4 \times k_BT_c$). In the clean limit there will
be large deviations  below the scaling line because $\rho_s$
remains constant while $\sigma(T_c) \times T_c$ can be extremely
large. Using data for Pb alloys\cite{Egloff} and for Nb
alloys\cite{DeSorbo} we find that, for the Pb sample on the Homes
scaling line, $\xi_0/\ell$ = 1.2 while the two Nb samples have
$\xi_0/\ell$ = 0.6 and 2.1. From eq. (2) it can be seen that in
the intermediate situation where $\xi_0/\ell \approx 1$ the above
factor of two is compensated. The apparent scaling here is
understandable but fortuitous.

Homes $et$ $al$\cite{Homes} consider Josephson coupling along the
$c$-axis and show the proportionality $\rho_s \propto \sigma
\times T_c$ but do not evaluate the proportionality constant. The
$c$-axis penetration depth, $\lambda_c$ is expressed in terms of
the Josephson current density, $J_c$ as $\lambda_c^2 =
\hbar/(2\mu_0deJ_c)$ where $d$ is the spacing between SC planes
and, for low temperatures, $J_c =
\pi\Delta/(2eR_n)$\cite{Ambegaokar}. $R_n = d/\sigma_c$ is the NS
tunnelling resistance between layers. For a $d$-wave gap we may
assume that the effective gap parameter, $\Delta_{eff} =
\Delta_0/\surd2$ where $\triangle_0$ ($\approx2.4 k_BT_c$) is the
maximum $d$-wave gap near $(\pi,0)$. Thus $\lambda_c^{-2} =
6.79(\pi^2/\hbar c^2)k_BT_c\sigma_c$ giving the relation
$\hbar/\tau = 2.23 \times k_BT_c$. If instead we consider a simple
rectangular I-V Josephson characteristic, then
$I_c=2\Delta_0/eR_n$ and the relation becomes $\hbar/\tau = 2.83
\times k_BT_c$, again in excellent agreement with the scaling
curve.

Thus, whether we consider Josephson coupling, dirty-limit SC or
marginal-Fermi-liquid SC we recover the observed scaling behavior.
Now we turn to the experimental data for the doping dependence of
the in-plane dynamics.

\begin{figure}
\centerline{\includegraphics*[width=95mm]{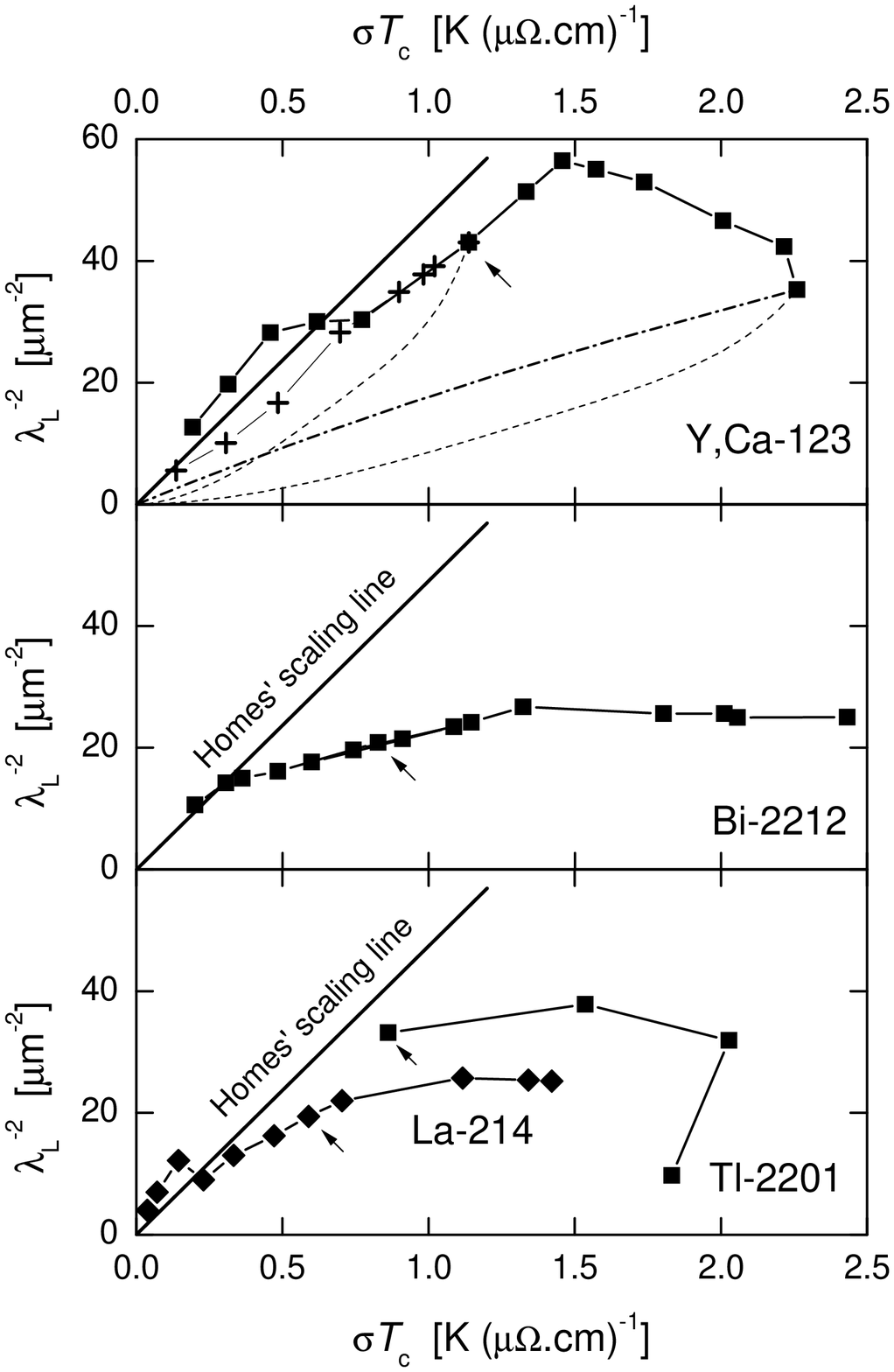}}
\caption{\small Superfluid density, $\rho_s$ versus $\sigma(T_c)
\times T_c$ for (a) Y$_{1-x}$Ca$_x$Ba$_2$Cu$_3$O$_{7-\delta}$, (b)
Bi$_2$Sr$_2$CaCu$_2$O$_{8+\delta}$ and (c)
Tl$_2$Ba$_2$CuO$_{6+\delta}$ (overdoped only) and
La$_{2-x}$Sr$_x$CuO$_4$. Arrows indicate optimal doping. In (a)
the effects of Zn-induced pairbreaking are shown (crosses) along
with the calculated A-G pairbreaking effects for constant $\sigma$
(dashed curve) and for $\sigma$ reduced by impurity scattering
(dash-dot curve). The Homes' data lie within $\sim\pm25\%$ of the
scaling line.}
\end{figure}

In Fig. 1 the filled squares and diamonds show values of
$\rho_s(0)$ plotted versus $\sigma(T_c)\times T_c$ for four
different HTS materials. The samples, technique for measuring
$\rho_s(0)$ and sample-type are as follows: (a)
Y$_{1-x}$Ca$_x$Ba$_2$Cu$_3$O$_{7-\delta}$ (transverse-field muon
spin relaxation - $\mu$SR, polycrystalline), (b)
Bi$_2$Sr$_2$CaCu$_2$O$_{8+\delta}$ (field-dependent specific heat,
polycrystalline), and (c) Tl$_2$Ba$_2$CuO$_{6+\delta}$ ($\mu$SR,
polycrystalline) with La$_{2-x}$Sr$_x$CuO$_4$ (ac-susceptibility,
oriented powders). The $\rho_s(0)$ data for
Tl$_2$Ba$_2$CuO$_{6+\delta}$ are from ref. \cite{Niedermayer} and
the remainder are from references in \cite{Tallon1}. The
techniques noted are well established except for the specific-heat
method. But $\rho_s(0)$ has also recently been measured for
Bi$_2$Sr$_2$CaCu$_2$O$_{8+\delta}$ using
ac-susceptibility\cite{Oh}. Apart from the heavily underdoped
region, the data are in excellent agreement with those used here
from specific-heat.

The in-plane $\sigma(T)$ data in Fig. 1 are from DC transport
measurements and for these it is essential to use single-crystal
measurements. These are available for all cited samples except
Y$_{1-x}$Ca$_x$Ba$_2$Cu$_3$O$_{7-\delta}$ for which we have used
thin-film data from Fisher\cite{Fisher}, supported by more recent
data from Naqib $et$ $al$.\cite{Naqib}. Sources for the remaining
samples are: Bi$_2$Sr$_2$CaCu$_2$O$_{8+\delta}$ from Watanabe $et$
$al.$\cite{Watanabe}; for Tl$_2$Ba$_2$CuO$_{6+\delta}$ from Tyler
and coworkers\cite{Tyler} and for La$_{2-x}$Sr$_x$CuO$_4$ from
Ando and coworkers\cite{Ando}. In all cases we evaluated
$\sigma(T_c)$ at the reported doping states, fitted to an
appropriate function of doping and interpolated to determine
$\sigma(T_c)$ at the doping state relevant to the available
$\rho_s$ data. For Bi$_2$Sr$_2$CaCu$_2$O$_{8+\delta}$, the
$\sigma(T_c)$ data had to be extrapolated to higher doping levels.
In Fig. 1 optimal doping (maximal $T_c$) is indicated by the
arrows and critical doping, where the pseudogap closes, is the
point of maximum $\rho_s$\cite{Tallon1}. The overdoped region lies
further to the right.

Our main conclusion is that each compound exhibits a large
departure from the ``universal" scaling line in the overdoped
region. In the optimal and underdoped regions, where we can
directly compare with Homes $et$ $al$, the situation is less
clear. In some cases our data differs markedly from theirs which
severely overestimates $\rho_s$ for Bi$_2$Sr$_2$CaCu$_2$O$_8$
relative to data from specific heat, ac susceptibility and $\mu$SR
and underestimates $\rho_s$ for Tl$_2$Ba$_2$CuO$_6$. For this
compound, and for La$_{2-x}$Sr$_x$CuO$_4$, their $\sigma(T_c)$
value is about half that observed from DC transport in single
crystals\cite{Tyler}. The source of these discrepancies may
possibly be attributable to the extrapolation to $\omega$ = 0 and
$\omega$ = $\infty$ required for the Kramers-Kronig transformation
of the optical data.

Bearing in mind these differences our underdoped results show a
consistent trend. For Bi$_2$Sr$_2$CaCu$_2$O$_8$,
Tl$_2$Ba$_2$CuO$_6$ and La$_{2-x}$Sr$_x$CuO$_4$ the scaling line
is not reached until near, or below, 1/8$^{th}$ doping. On the
other hand Y$_{1-x}$Ca$_x$Ba$_2$Cu$_3$O$_7$, though not linear,
roughly follows the trend of the scaling line across the entire
underdoped region. However, there are two reasons why this could
misrepresent the situation. Firstly, compared with other bilayer
cuprates with similar $T_c$ this compound generally exhibits a
higher superfluid density arising from the contribution of the
chain layer. This may be approximately balanced in the scaling
plot by the additional conductivity also arising from the chains
and so we put this effect to one side. But secondly, the
conductivity here is for thin films and thus is probably
diminished relative to single-crystal data. All the data points
may lie further to the right, perhaps by a factor of the order of
1.5. Then the behavior would be rather similar to that shown by
Bi$_2$Sr$_2$CaCu$_2$O$_{8+\delta}$ and La$_{2-x}$Sr$_x$CuO$_4$. It
may be that there is no correlation with the scaling line until
the lowest doping state around $p$ = 1/8 and lower.

What can we conclude from this? One implication may be that in
this heavily underdoped region the in-plane physics is rather
similar to the out-of-plane physics, namely that of
Josephson-coupled SC domains. Here the electronic state is
probably granular, with SC patches separated by insulating
Josephson barriers. Such a model has been proposed even for
optimally doped cuprates and STM suggests the presence of
nanoscale inhomogeneity\cite{Pan,Lang}. However this has been
questioned\cite{Bobroff}, and in particular the electronic state
seen from the perspective of NMR and specific heat seems to be
homogeneous for $p > 0.12$\cite{Loram}. Below that doping state
these techniques reveal the presence of normal
quasiparticles\cite{Benseman}. Here the electronic state
definitely appears to become spatially heterogeneous.

We turn now to ask the question as to whether the universal
scaling relation may also apply in the case of impurity
scattering. This behavior can be understood in terms of
generally-accepted theories for the pair-breaking effects of Zn
impurities, namely the suppression of $T_c$ by non-magnetic
scattering, which for a $d$-wave superconductor suppresses $T_c$
according to the Abrikosov-Gor'kov (A-G) formula and $\rho_s$
according to the work of Maki and coworkers\cite{Maki}. When we
add in the effects of scattering on $\sigma(T)$ (described by
Matthiessen's rule) this roughly reproduces the observed scaling
behavior for optimally-doped samples but, again, we expect large
deviations for overdoped samples for which conductivity data as a
function of Zn doping are not yet available. The theory can be
developed along the following lines.

For a non-magnetic scatterer $d$-wave SC is rapidly suppressed and
the effects on $T_c$, $\rho_s$ and $\sigma$ as a function of
impurity concentration, $y$, are governed by the density of states
(DOS) at the Fermi level. These effects have been shown to be
fully consistent with thermodynamic data\cite{Tallon2}. The
reduction in $T_c$ is given by:
\begin{equation}
\ -\ln(T_c/T_{c0}) = \psi[1/2 + \Gamma/\Gamma_c] - \psi[1/2],
\end{equation}
where $\psi$[$x$] is the digamma function, $T_{c0}$ = $T_c(y$=0).
For unitary scattering, $\Gamma = n_i/\pi g(E_F)$ is the
pair-breaking scattering rate and $\Gamma_c$ is its critical value
for which $T_c$ is suppressed to zero. Here $g(E_F)$ is the DOS
per spin, $n_i (=\alpha y/abc)$ is the density of impurities,
$\alpha$ is the number of CuO$_2$ planes per unit cell and $a$,
$b$ and $c$ are the lattice parameters. As before\cite{Tallon2},
we adopt the strong-coupling scenario $\Gamma_c = 1.65k_BT_{c0}$
and determine $g(E_F)$ from the electronic specific heat
coefficient, $\gamma$, using $\gamma = (2/3)\pi^2k_B^2g(E_F)$. The
following linearized form of the A-G equation is valid up to about
(2/3)$\Gamma_c$:
\begin{equation}
\ T_c/T_{c0} = 1 - 0.69\Gamma/\Gamma_c = 1 - (0.86\alpha R/\gamma
T_{c0})\times y ,
\end{equation}

\noindent where $R = N_Ak_B$. In the overdoped region $\gamma$ is
constant, independent of temperature and doping, so that the
slope, $\partial T_c/\partial y$, in eqns. (3) or (4) remains
constant. As a practical measure\cite{Tallon2}, the value of
$\gamma$ was taken as $S/T$ evaluated at $T_c$, where S is the
electronic entropy. This is just the average of $\gamma(T)$
between $T=0$ and $T_c$.

The calculation of the reduction in superfluid density is more
complex and is carried out numerically\cite{Maki}. However, we
find an excellent approximation to within a few percent across the
entire pair-breaking range using:
\begin{equation}
\ \rho_s/\rho_{s0} \approx {{T_c/T_{c0}}\over{1 + 1.7(1 -
T_c/T_{c0})}}.
\end{equation}

This non-linear relation reflects the fact that $\rho_s$ is
initially suppressed much faster than $T_c$\cite{Tallon1}. Eqns.
(3) and (5) are universal relations and if impurity scattering
were to have no effect on $\sigma$ then the effect of Zn
substitution would be given by the dashed A-G curves shown in Fig.
1(a) for the optimal and most overdoped
Y$_{1-x}$Ca$_x$Ba$_2$Cu$_3$O$_{7-\delta}$.

In fact, the resistivity is increased by impurity scattering
according to a linear Matthiessen law. The resultant decrease in
$\sigma(T)$ seems to effectively linearize the simple A-G curve in
the scaling plot. We illustrate this using the experimental data
shown by the crosses in Fig. 1. These are data for an
optimally-doped sample where the reduction in $\rho_s$ and $T_c$
is from $\mu$SR measurements by Bernhard $et$
$al$.\cite{Bernhard}, and the $\sigma(T)$ data for Zn substitution
at the same doping state are from thin-film samples studied by
Naqib $et$ $al$\cite{Naqib}. These trend back linearly to the
origin in reasonably good agreement with the scaling curve. We now
seek to justify this quantitatively.

The residual planar resistivity arising from impurity scattering
is given by\cite{Fukuzumi} 4($\hbar/e^2)(n_i/n)sin^2\delta_0$
where $\delta_0$ is the $s$-wave scattering phase shift which, in
the unitary limit, we take to be $\pi/2$. Thus the total
resistivity is $m^*/(ne^2\tau) + 4c(\hbar/\alpha e^2)(n_i/n)$ and
$\sigma$ becomes
\begin{equation}
\ \sigma =  {{ne^2\tau/m^*}\over{1 + (4\hbar\tau/abm^*)\times y}}
\end{equation}
where $\tau$ is evaluated at $T_c(y)$.

There are two impurity effects to be considered. Firstly, there is
the Matthiessen term in the denominator of eqn.(6) and, secondly,
$\sigma(T)$ must be evaluated $at$ $T_c$ which is itself reduced
by scattering. This second effect is accommodated by the fact
that, in eqn. (7) below, the only two variables in the
coefficients are $\tau$ and $T_c$ and they appear as the product
$\tau T_c$. Thus, provided the material lies on the scaling curve
in the absence of impurities, this product is a constant.
Proceeding, the explicit variable $y$ in eqn. (6) may be
eliminated by substituting from eqn. (4) to give:
\begin{equation}
\ \sigma(y)T_c(y) = {{\sigma(0)T_{c0}\times(T_c(y)/T_{c0})}\over{1
+ (\beta -1)(1 - T_c(y)/T_{c0})}}.
\end{equation}
where $\beta = 4\hbar\gamma\tau_0 T_{c0}/0.86\alpha Rabm^*$ and
$\tau_0=\tau$ evaluated at $T_c(y=0)$. Taking $m^* \approx 2m_e$,
as observed for nodal quasiparticles\cite{Kim}, we have only to
note that the right side of eqn.(7) is of the same form as eqn.(5)
with the coefficient $(\beta -1) = 1.9$. Thus
\begin{equation}
\ \sigma(y)T_c(y)/\rho_s(y) \approx \sigma(0)T_{c0}/\rho_{s0},
\end{equation}
and the impurity scattering data remain on the scaling curve if,
in the first instance, it lies on the curve in the absence of
impurity scattering. This seems to explain the crosses shown in
Fig. 1. This calculation shows that, with underdoping, the curve
would steepen due to the further opening of the pseudogap (and the
associated reduced entropy at $T_c$) while with overdoping it
would flatten (due to the closing of the pseudogap). It would be
nice to test this but we do not have combined $\rho_s(y)$ and
$\sigma(y)$ data as a function of Zn concentration for heavily
overdoped samples. Nevertheless, we have calculated the expected
effect of Zn substitution using eqn. (7) and this is plotted in
Fig. 1 by the dotted line. This shows a fairly linear behavior but
which is displaced far from the ``universal" scaling curve.

In conclusion, we have examined the scaling relation, $\rho_s
\propto \sigma(T_c) \times T_c$, and find this is equivalent to
the relation $\hbar/\tau = 2.7\pm0.5 \times k_BT_c$ which could
equally arise from (i) conventional dirty-limit SC, (ii)
marginal-Fermi-liquid behaviour, (iii) Josephson-coupling along
the $c$-axis of the CuO$_2$ planes, or (iv) unitary-limit impurity
scattering. We conclude that the Pb and Nb samples correlate
because these samples are near the dirty limit. Interestingly, the
Homes scaling line corresponds to the mean free path just above
$T_c$ being approximately $2\times\xi_0$ (using the maximum gap
$\approx 2.38k_BT_c$ for a weak-coupling $d$-wave SC). Even more
intriguingly, in terms of standard microscopic BCS
theory\cite{Waldram}, it corresponds to the Gor'kov kernel having
a range $\xi_G$ near $T_c$ equal to the mean free path just above
$T_c$.

When we examine the doping dependence of the in-plane $\rho_s$ and
$\sigma(T_c) \times T_c$ we find a total breakdown of the scaling
relation across the overdoped region. This shows that the mean
free path at $T_c$ is now much larger than both of the above-noted
coherence lengths. This breakdown may extend into the underdoped
region, allowing for the fact that the only exception is for
$thin$-$film$ Y$_{1-x}$Ca$_x$Ba$_2$Cu$_3$O$_{7-\delta}$. The
recovery of the scaling behavior at very low doping may arise from
similar physics to $c$-axis Josephson coupling, namely,
weak-linked patches of superconducting domains in the
inhomogeneous strongly-underdoped state.

%%\end{multicols}
\end{document}